\documentclass[letter,traditabstract,longauth]{aa} 
\usepackage{graphicx}
\usepackage{txfonts}
\usepackage{natbib}
\bibpunct{(}{)}{;}{a}{}{,} 

\begin{document}


\title{Herschel images of NGC 6720: H$_2$ formation on dust grains\thanks{Herschel is an ESA space observatory with science instruments provided by European-led Principal Investigator consortia and with important participation from NASA.}}

\author{P.A.M. van Hoof\inst{1}\thanks{email: p.vanhoof@oma.be}
  \and
  G.C. Van de Steene\inst{1}
  \and
  M.J. Barlow\inst{2}
  \and
  K.M. Exter\inst{3}
  \and
  B. Sibthorpe\inst{4}
  \and
  T.~Ueta\inst{5}
  \and
  V.~Peris\inst{12}
  \and
  M.A.T.~Groenewegen\inst{1}
  \and
  J.A.D.L.~Blommaert\inst{3}
  \and
  M.~Cohen\inst{6}
  \and
  W.~De~Meester\inst{3}
  \and
  G.J.~Ferland\inst{7}
  \and
  W.K.~Gear\inst{8}
  \and
  H.L.~Gomez\inst{8}
  \and
  P.C.~Hargrave\inst{8}
  \and
  E.~Huygen\inst{3}
  \and
  R.J.~Ivison\inst{4}
  \and
  C.~Jean\inst{3}
  \and
  S.J.~Leeks\inst{9}
  \and
  T.L.~Lim\inst{9}
  \and
  G.~Olofsson\inst{10}
  \and
  E.T.~Polehampton\inst{9,11}
  \and
  S.~Regibo\inst{3}
  \and
  P.~Royer\inst{3}
  \and
  B.M.~Swinyard\inst{9}
  \and
  B.~Vandenbussche\inst{3}
  \and
  H.~Van~Winckel\inst{3}
  \and
  C.~Waelkens\inst{3}
  \and
  H.J.~Walker\inst{9}
  \and
  R.~Wesson\inst{2}
}

\authorrunning{van Hoof et al.}

\institute{Royal Observatory of Belgium, Ringlaan 3, B-1180 Brussels, Bel\-gium
  \and
  Dept of Physics \& Astronomy, University College London, Gower St, London WC1E 6BT, UK 
  \and
  Instituut voor Sterrenkunde, Katholieke Universiteit Leuven, Ce\-les\-tij\-nenlaan 200 D, B-3001 Leuven, Belgium
  \and
  UK Astronomy Technology Centre, Royal Observatory Edinburgh, Blackford Hill, Edinburgh EH9 3HJ, UK
  \and
  Dept. of Physics and Astronomy, University of Denver, Mail Stop 6900, Denver, CO 80208, USA 
  \and
  Radio Astronomy Laboratory, University of California at Berkeley, CA 94720, USA
  \and
  University of Kentucky, Dept.\ of Physics and Astronomy, 177 CP Building, Lexington, KY 40506--0055, USA
  \and
  School of Physics and Astronomy, Cardiff University, 5 The Parade, Cardiff, Wales CF24 3YB, UK
  \and
  Space Science and Technology Department, Rutherford Appleton Laboratory, Oxfordshire, OX11 0QX, UK  
  \and
  Dept.\ of Astronomy, Stockholm University, AlbaNova University Center, Roslagstullsbacken 21, 10691 Stockholm, Sweden 
  \and
  Department of Physics, University of Lethbridge, Lethbridge, Alberta, T1J 1B1, Canada 
  \and
  Astronomical Observatory, Valencia University, Edifici Instituts d'Investigaci\'o, Parc Cient\'\i fic,
  C/ Catedr\'atico Agust\'\i n Escardino 7, 46980 Paterna (Valencia), Spain
}

\date{Received; accepted}

\abstract{Herschel PACS and SPIRE images have been obtained of NGC 6720 (the
  Ring Nebula). This is an evolved planetary nebula with a central star that
  is currently on the cooling track, due to which the outer parts of the
  nebula are recombining. From the PACS and SPIRE images we conclude that
  there is a striking resemblance between the dust distribution and the H$_2$
  emission, which appears to be observational evidence that H$_2$ forms on
  grain surfaces. We have developed a photoionization model of the nebula with
  the Cloudy code which we used to determine the physical conditions of the
  dust and investigate possible formation scenarios for the H$_2$. We conclude
  that the most plausible scenario is that the H$_2$ resides in high density
  knots which were formed after the recombination of the gas started when the
  central star entered the cooling track. Hydrodynamical instabilities due to
  the unusually low temperature of the recombining gas are proposed as a
  mechanism for forming the knots. H$_2$ formation in the knots is expected to
  be substantial after the central star underwent a strong drop in luminosity
  about one to two thousand years ago, and may still be ongoing at this
  moment, depending on the density of the knots and the properties of the
  grains in the knots.}


\maketitle

\section{Introduction}

Grains play an important role in many environments, including planetary
nebulae (PNe), because of extinction, photoelectric heating, their influence
on the charge and ionization balance of the gas, as catalysts for
grain-surface chemical reactions (e.g.\ H$_2$ formation), and as seeds for
freeze-out of molecules. Previous satellite missions such as IRAS, ISO,
Spitzer, and AKARI have allowed us to study the dust in PNe, but unfortunately
the angular resolution of these instruments was too low to get detailed
information on the spatial distribution of the dust. This has now changed with
the launch of Herschel, which allows us to study the spatial structures in
unprecedented detail. In this paper we will do this for NGC 6720 (M57, the
Ring nebula) to study H$_2$ formation. NGC 6720 is an evolved, oxygen-rich
bipolar nebula seen nearly pole-on. The nebula is optically thick to ionizing
radiation in most directions, but optically thin in the polar regions
\citep[hereafter OD07]{OD07}. Molecules such as H$_2$ and CO have been
detected \citep{Be78, Hu86}. The central star has exhausted hydrogen shell
burning and is now on the cooling track. As a result the outer halo is
recombining and re-ionization of the innermost recombined material due to
expansion of the nebula has just started (OD07). This object is very similar
to the Helix nebula, which seems to be further advanced along the same
evolutionary path (OD07). In Sect.~\ref{obs} we will describe the
observations, and in Sect.~\ref{origin} we will discuss various scenarios for
the formation of H$_2$.

\section{Observations}
\label{obs}

The images of NGC 6720 presented in this paper were obtained with the PACS and
SPIRE instruments on board the Herschel satellite \citep{Pi10} on 2009-10-10
and 2009-10-06, respectively, as part of the Science Demonstration Phase of
the Mass-loss of Evolved StarS (MESS) guaranteed time key program (Groenewegen
et al., in preparation). The PACS instrument is described by \citet{Po10}. The
SPIRE instrument, its in-orbit performance, and its scientific capabilities
are described by \citet{Gr10}, and the SPIRE astronomical calibration methods
and accuracy are outlined by \citet{Sw10}. The reduction of our data is
described in Appendix~\ref{reduction}. We also determined the total flux in
each of the images. This is described in Appendix~\ref{photometry}. Below we
will discuss these images in detail and compare them to the H$_2$ morphology
of the nebula. To this end we used a ground-based H$_2$ 2.12~$\mu$m image,
obtained with the Omega2000 camera on the 3.5-m Zeiss telescope at the Calar
Alto Observatory.

\begin{figure*}
\includegraphics[width=1.00\textwidth]{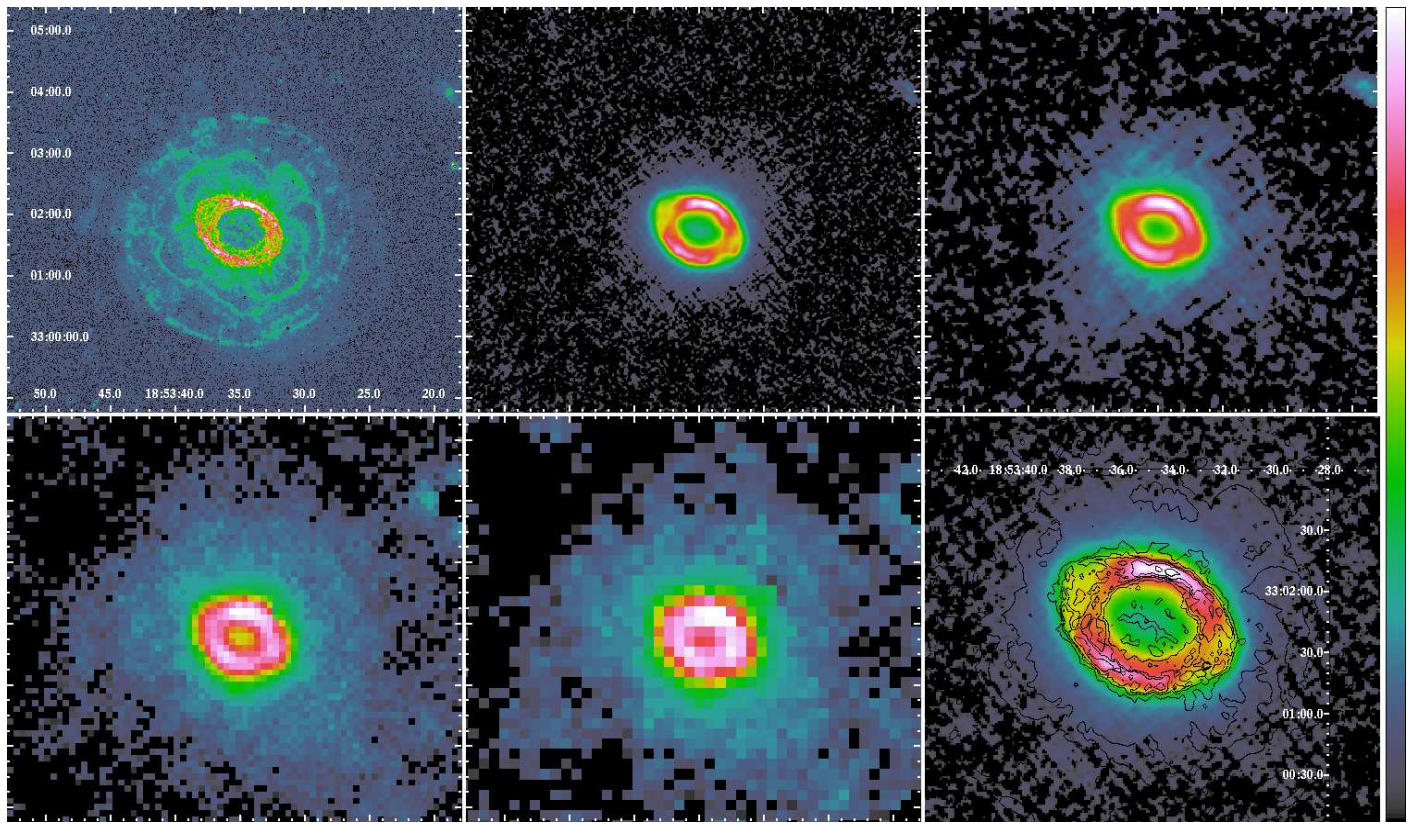}
\caption{NGC 6720 in five photometric bands. Top row from left to
  right: H$_2$ 2.12~$\mu$m, PACS 70~$\mu$m, PACS 160~$\mu$m. Bottom row from
  left to right: SPIRE 250~$\mu$m, SPIRE 350~$\mu$m, and an overlay of the
  Calar Alto H$_2$ contours on the PACS 70~$\mu$m image.
  The H$_2$ image is not flux calibrated. The maps have standard orientation
  (N to the top, E to the left).}
\label{composite}
\end{figure*}

In Fig.~\ref{composite} we present the PACS and SPIRE images and compare them
to the Calar Alto H$_2$ image. The morphology of the dust and the H$_2$
emission shows a striking resemblance even in small details. This appears to
be observational evidence that H$_2$ forms on grain surfaces in an
astrophysical environment. Such evidence is very rare. A similar result was
obtained by \cite{Ha03} for the $\rho$ Ophiuchi molecular cloud, suggesting
that H$_2$ forms on PAH surfaces. The presence of PAHs in NGC 6720 cannot be
fully excluded, but seems very unlikely. Accordingly our observations are the
first indication for H$_2$ formation on oxygen-rich dust grains in an
astrophysical environment to our knowledge.

We can divide the H$_2$ image into 3 regions: the inner ring, with a semimajor
axis of $\sim45$\arcsec, an inner set of arcs with a radius of $\sim70$\arcsec\
(the inner halo), and an outer set of arcs with a radius of $\sim110$\arcsec\
(the outer halo). There is also fainter emission outside of these outer arcs,
but this is not very easy to see on the printed map.

Comparing the H$_2$ and the sub-mm maps it is clear that the size and shape of
the inner ring is reproduced on all images. The same can be said for the inner
halo, which can be seen clearly on all maps. In the NW part of the inner halo,
the H$_2$ emission region is broader than in the SE---this too is the case on
the SPIRE images, and less clearly so on the PACS images. Finally, the
outer halo on the H$_2$ image is also visible on the sub-mm images---as
extended emission on the SPIRE images and as a faint circular ring shape on
the PACS images (on the $70$\,$\mu$m image most clearly). On the zoomed PACS
$70$\,$\mu$m with overlaid H$_2$ contours (Fig.~\ref{composite}, lower right
panel) the clear correspondence between the optical and sub-mm emission in the
ring and the inner halo is highlighted.

\begin{table}
\caption{Parameters of the Cloudy model of NGC~6720.}
\label{model:par}
\begin{tabular}{lrlr}
\hline
$T_{\rm eff}$ (kK)            & 134.7    & $\epsilon$(C)                 &  8.20    \\
$L_{\ast}$ (L$_{\odot}$)      &  239.    & $\epsilon$(N)                 &  7.89    \\
$r_{\rm in}$ (mpc)            & 56.      & $\epsilon$(O)                 &  8.50    \\
$r_{\rm out}$ (mpc)           & 175.     & $\epsilon$(Ne)                &  7.93    \\
log($n_{\rm H})$ (cm$^{-3}$)  & 2.60     & $\epsilon$(S)                 &  6.42    \\
$T_{\rm e}$ (kK)              & 11.97    & $\epsilon$(Cl)                &  4.94    \\
log($n_{\rm e})$ (cm$^{-3}$)  & 2.62     & $\epsilon$(Ar)                &  6.19    \\
log($\Gamma$)                 & $-$2.21  & $\epsilon$(Fe)                &  5.18    \\
$\epsilon$(He)                & 11.00    & $D$ (pc)                      &  740.    \\
\hline
\end{tabular}
\end{table}

\section{The origin of the H$_2$}
\label{origin}

From the ground-based H$_2$ images it is clear that most of the H$_2$ resides
in high density knots in the inner ring \citep{Sp03}. In this section we will
investigate the origin of this H$_2$. The halo also shows H$_2$ emission,
which has a very different morphology though. The latter will not be discussed
here. It is clear that H$_2$ was formed in the dense AGB wind. In the post-AGB
phase three different scenarios will be investigated: 1) this H$_2$ survived
in the ionized region, 2) this H$_2$ survived in the knots, which formed
before the gas was ionized and 3) this H$_2$ was destroyed and then was formed
again later inside the knots when they formed. These scenarios were already
investigated for the Helix nebula by \citet{Ma09}. In order to determine the
physical conditions in the nebula we created a photoionization model using a
prerelease of version C10.00 of the photoionization code Cloudy (revision
3862), last described by \citet{Fe98}. The details of the modeling are
described in Appendix~\ref{model}, the resulting model parameters are
summarized in Table~\ref{model:par}.

\subsection{Can H$_2$ survive in the ionized region?}
\label{survive}

For the Helix nebula, \citet{Ma09} conclude that ``part of the H$_2$ is
primordial, i.e. formed during the AGB phase, and survived the ionization of
the nebula''. This conclusion is based on the work of \citet[hereafter
AG04]{AG04}. We recreated the models of AG04 with Cloudy. This code is capable
of creating an equilibrium model of the ionized region as well as the PDR and
the molecular regions. It includes state-of-the-art code for modeling grains
\citep{vH04} as well as the chemistry network \citep{Ab05} and the H$_2$
molecule \citep{Sh05}. We created spherically symmetric, one-dimensional
models with the parameters of the standard model of AG04. Our H$_2$ model
includes all ro-vibrational levels of the ground electronic state, as well as
the six electronically excited states that are coupled to the ground state by
permitted electronic transitions. Self-shielding of the H$_2$ molecule was
calculated self-consistently using detailed radiative transfer (including line
overlap) of each of the roughly half a million lines of the molecule.

\begin{figure}
\includegraphics[width=0.9\columnwidth]{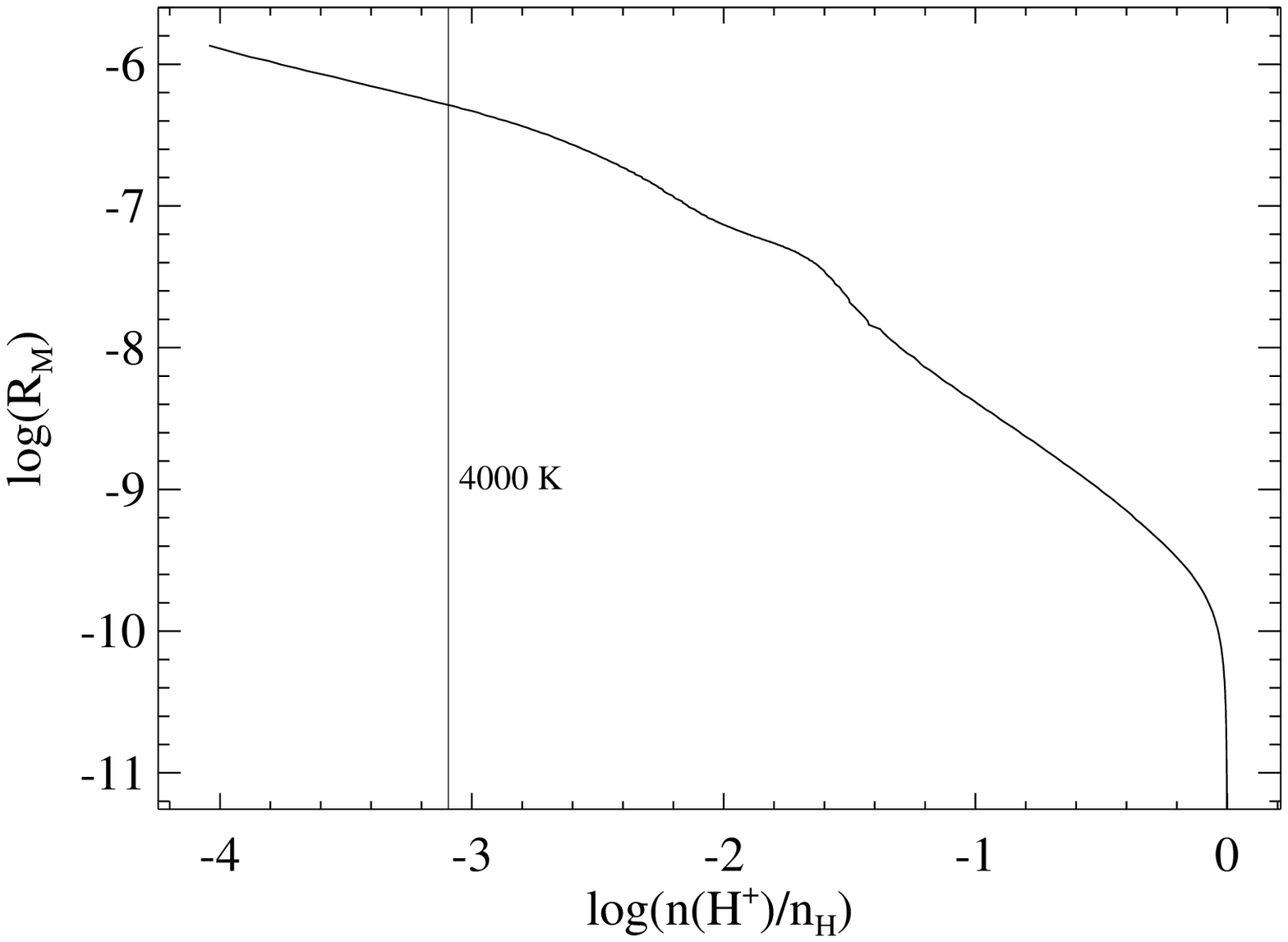}
\caption{The H$_2$ mass fraction as a function of the stopping criterion.
The vertical line indicates where $T_{\rm e}$ reaches 4000~K.}
\label{RMvsstop}
\end{figure}

\begin{figure}
\includegraphics[width=0.9\columnwidth]{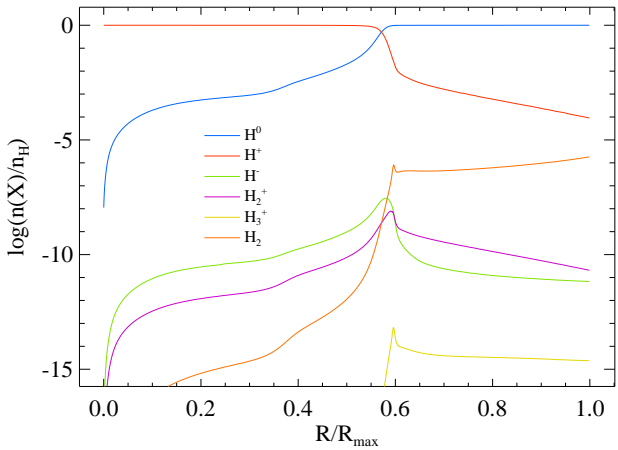}
\caption{The standard model of AG04 recreated with Cloudy.}
\label{standmod}
\end{figure}

\begin{figure}
\includegraphics[width=0.9\columnwidth]{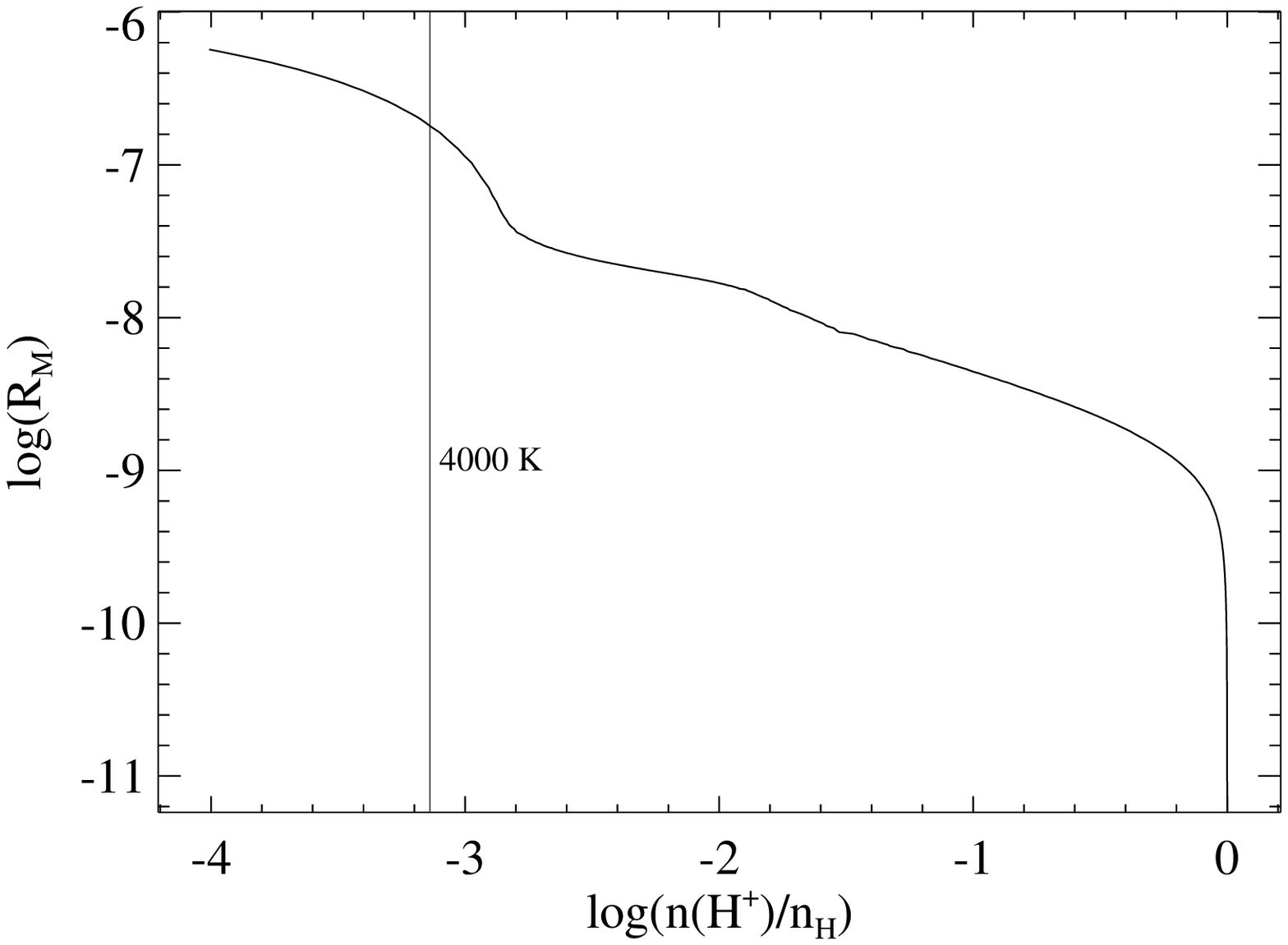}
\caption{Same as Fig.~\ref{RMvsstop}, but for the standard4 model.}
\label{RMvsstop4}
\end{figure}

\begin{figure}
\includegraphics[width=0.9\columnwidth]{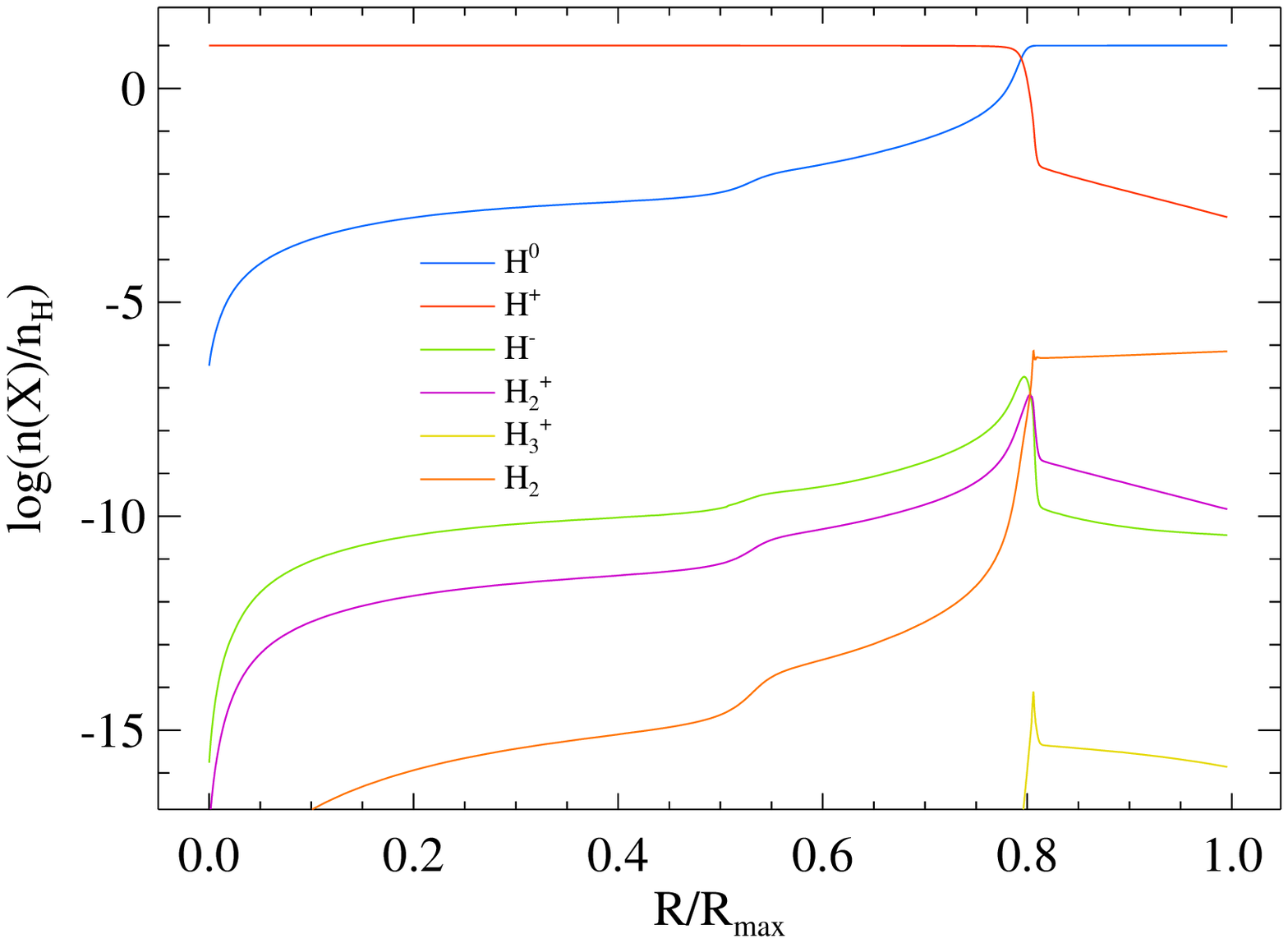}
\caption{Same as Fig.~\ref{standmod}, but for the standard4 model.}
\label{standmod4}
\end{figure}

First we needed to define the outer radius of the ionized region. In PNe that is
not straightforward. Due to the $\nu^{-3}$ dependence of photoionization cross
section, the high energy tail of the stellar spectrum will propagate through
the ionized region into the PDR and cause hydrogen to stay partially ionized
there. Such PDRs are sometimes referred to as XDRs. The transition from fully
ionized gas to neutral gas will be gradual and the transition layer will have
a considerable thickness. \citet{AG04} stopped their calculations when $n({\rm
  H}^+)/n_{\rm H} = 10^{-4}$ was reached\footnote{The number $10^{-4}$ is
  stated in the text, but an inspection of their Fig.~1 suggests that they
  actually used $10^{-3}$. In this section we will adopt $10^{-4}$ as the
  stopping criterion.}. Cloudy by default stops when the electron temperature
drops below 4000~K, based on the notion that at lower temperatures the
forbidden optical emission lines will no longer be excited effectively and the
stopping radius will coincide with the outer radius in optical images of the
nebula. The choice of the stopping criterion is crucial, but is unfortunately
not discussed by AG04. The H$_2$ fraction $R_{\rm M}$ (Eq.~7 in AG04)
continually rises as the value of $n({\rm H}^+)/n_{\rm H}$ where the
calculation is stopped, is lower. This is illustrated in Fig.~\ref{RMvsstop}.
This clearly shows that there is no obvious choice for the stopping criterion:
the deeper one integrates, the higher the value of $R_{\rm M}$ becomes. This is
hardly surprising: deeper layers are better shielded from UV radiation and the
H$_2$ abundance will be higher. The local maximum of the H$_2$ abundance that
AG04 report around $n({\rm H}^+)/n_{\rm H} = 0.1$ is barely reproduced in the
Cloudy model and has virtually no effect on the overall value of $R_{\rm M}$
(see Fig.~\ref{standmod}). Most of the H$_2$ present in this model will be in
regions near the outer radius. The H$_2$ weighted average ionization of
hydrogen is slightly more than $10^{-3}$. So a more precise statement would be
that a small fraction of H$_2$ can survive (or better: is constantly destroyed
and formed again) in the transition region between the ionized region and the
molecular region. The expectation of AG04 that using a detailed treatment of
H$_2$ self-shielding could increase the value of $R_{\rm M}$ by up to 2 dex
could not be confirmed with Cloudy, at least not for the standard model of
AG04. Our value of $R_{\rm M}$ is actually slightly lower than reported by
AG04. The standard model of AG04 is roughly appropriate for a low-mass central
star on the horizontal track. To model a high-mass central star, we also
created a ``standard4'' model which is identical to the standard model except
that $n_H = 10^4$~cm$^{-3}$ and $L_{\ast}$ = $10^4$~L$_{\odot}$. The results
are shown in Figs.~\ref{RMvsstop4} and \ref{standmod4}. In this model, the
value for $R_{\rm M}$ is somewhat lower than in the standard model. So the
conclusion is that a small amount H$_2$ can exist in the ionized region,
mostly in the transition zone towards the PDR, and that this amount is not
sufficient to explain the observed H$_2$ emission in the inner ring.

\subsection{Can H$_2$ survive in knots?}

When the photoionization of the AGB shell starts at the moment when a
planetary nebula is born, the molecules in the circumstellar shell will be
very quickly destroyed unless they can somehow be shielded from the ionizing
radiation. Dense knots that formed during the AGB phase can provide this
environment and seem a very attractive explanation for the existence of H$_2$
in evolved PNe such as NGC 6720. It is clear that dense knots can shield the
material in the center from FUV flux and allow H$_2$ molecules to survive. But
it is also obvious from the images presented in e.g.\ \citet{Ma09} that the
FUV flux will interact with the outer layers of the knot. This interaction
causes material to be advected off the knot. During that process the material
will be heated, causing H$_2$ to shine. This provides a natural explanation
for the H$_2$ spectrum seen in the Helix nebula \citep{He07}. This process
will gradually erode the knots, and eventually they will be destroyed. We will
now make an estimate of the lifetime of a knot in the ionized region.

We adopt a value of 1.5$\times 10^{-5}$~M$_\odot$ for the mass of the knots
\citep{Me05, Ma09}. We furthermore assume that the knots have a typical
diameter of 0.5\arcsec. Combined with a distance of 740~pc \citep{OD09} this
gives a physical diameter of 5.54$\times10^{10}$~km (370~AU), in agreement
with the typical size stated in \citet{OD03}. Assuming that 70\% of the mass
is hydrogen, this yields an average number density of hydrogen of $n_{\rm H} =
$1.4$\times 10^{5}$~cm$^{-3}$. The density in the center of the knot should be
higher, in agreement with the measurements by \citet{Me98} and \citet{Hu02}.
We assume spherical astronomical silicate grains to be present in the knot,
with a standard ISM size distribution \citep{MRN} and a dust-to-gas mass ratio
equal to the standard model of NGC~6720.

\citet{He07} state that their models C06 and A06 matched the observations of
the Helix nebula best for the outer and inner knots, respectively.
Taking the parameters for the advection flow presented in Table~1 of
\citet{He07}, one can calculate that the erosion rate would be approximately
10$^{-10}$ and 10$^{-9}$~M$_\odot$\,yr$^{-1}$ for the C06 and A06 model,
respectively. Combining this with the adopted mass of the knot yields a
survival time of 15,000 to 150,000 yr. However, one should realize that during
most of PN evolution the central star luminosity was much higher than assumed
here (by upto 2 dex, the central star of the Helix nebula has a luminosity of
120~L$_\odot$), and also that the nebula was much more compact during the
early stages of evolution causing the knots to be much closer to the central
star. The will cause the erosion to be much faster early on and hence the
survival time to be much shorter. The erosion rate will scale at least with
the square-root of the EUV flux when the advection flow is in the
recombination-dominated regime. So a conservative estimate is that the
survival time is at least a factor 10 shorter than stated above, and likely
more. \cite{OD07, OD09} state that the kinematic age of the Ring nebula is
7000~yr, making the survival of the knots inside the ionized region
problematic. Note that this argument also applies to models where the knots
formed due to instabilities at the ionization front during the onset of
ionization. More detailed modeling of the knots during the high-luminosity
phase is warranted though to reach a more definitive conclusion.

\subsection{Can H$_2$ be formed again after ionization?}

\citet{OD07} argue that knot formation only starts after the central star has
entered the cooling track and the nebular material starts recombining. This
argument is based on the fact that no similar features are seen in HST images
of younger PNe. We propose a new explanation in which the recombining gas
cools very quickly, much faster than the recombination proceeds. This is
indicated by Fig.~4 in \citet{Fe81}, but these calculations clearly need to be
redone for conditions that are more appropriate for NGC 6720. The fast cooling
would result in gas that is still ionized (and thus produces abundant
recombination radiation) but has very little thermal pressure support. The
radiation pressure of the recombination radiation on the dust and/or gas may
then cause the medium to become unstable and fragment into many globules.

\citet{OD07} also argued that the main gas phase of NGC 6720 was fully ionized
during the high-luminosity phase, so that any H$_2$ present in the gas would
have been destroyed very quickly (on timescales less than a year).
Consequently any H$_2$ present in the knots must have formed after the
formation of the knots itself started and they became dense enough to shield
the molecules from the UV light. To investigate this scenario further, we
created Cloudy models of the knots using revision 3862 of the experimental
``newmole'' version of the code in order to cope with the very steep density
gradient at the outer edge of the knot. As was the case with the models
described in Sect.~\ref{survive}, we use a full and self-consistent H$_2$
model. The models are based on our standard model of NGC~6720, with a modified
density law that mimics a dense knot just outside the ionized region. The
density is constant until the electron temperature drops below 4000~K, then
the density will rise steeply towards the core density of the knot with an
assumed scale-length of 2$\times 10^9$~km.

\begin{figure}
\includegraphics[width=0.9\columnwidth]{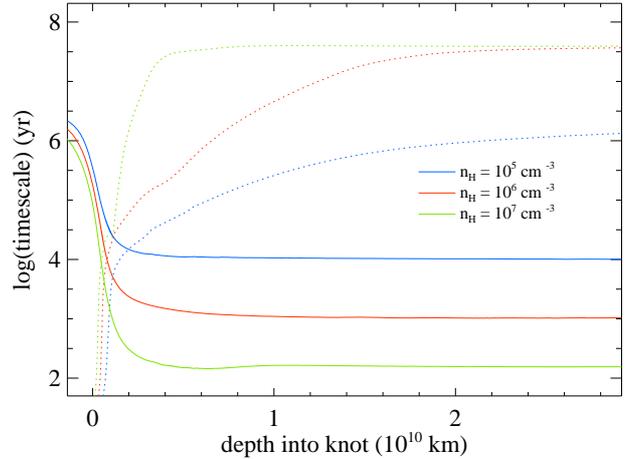}
\caption{The H$_2$ formation and destruction timescales (solid and dotted
  lines, respectively) for knots with a central density of 10$^5$, 10$^6$, and
  10$^7$~cm$^{-3}$ just outside the ionized region.}
\label{h2:timescales}
\end{figure}

We computed eight Cloudy models with $\log n_{\rm H} = 4\,(0.5)\,7.5$ to
cover the full range of plausible densities in the center of the knot. From
the models we extracted the formation (solid lines) and destruction timescales
(dotted lines) of H$_2$. The results are shown in Fig.~\ref{h2:timescales} for
selected models. The Cloudy models show that a minimum core density of roughly
10$^5$~cm$^{-3}$ is needed to get sufficient shielding from the UV radiation
for substantial H$_2$ formation in the knot. \citet{Go63} have shown that
H$_2$ formation is dominated by grain surface reactions for conditions similar
to the knots. The Cloudy models confirm this. The grain surface reaction rate
used in Cloudy is taken from \citet{CT02, CT02a}. The grain surface formation
rate scales linearly with the integrated projected grain surface area. The
latter depends critically on the assumed grain size distribution as well as
the shape of the grains and the grain abundance. The integrated grain surface
area in our models is $\Sigma = 9.78\times 10^{-22}$~cm$^{2}$/H. Using other
size distributions found in the literature \citep[e.g.][]{WD01} yielded
values for $\Sigma$ with a range of a factor of 5, with our adopted value
being towards the upper end of that range. If we assume that the grains have a
fractal shape rather than spherical, the projected grain surface area will be
larger than assumed above, yielding shorter formation timescales.

The formation timescales from the Cloudy models can be summarized in the
fitting formula given in Eq.~\ref{fitform}

\begin{equation}
\label{fitform}
t_{\rm H_2} = \frac{1170}{n_6^{0.9}\,\Sigma_{\,-21}}\ {\rm [yr].}
\end{equation}

Here $n_6$ is the hydrogen density in the knot in units of $10^6$~cm$^{-3}$
and $\Sigma_{\,-21}$ is the projected grain surface area in units of
$10^{-21}$~cm$^2$/H. \citet{OD07} state that the central star has exhausted
hydrogen-shell nuclear burning one to two thousand years ago. Given the
observed core densities in the knots, significant H$_2$ formation would be
possible (though conversion from H$^0$ to H$_2$ need not be complete) assuming
that the knot formation was quick after the recombination started and keeping
the uncertainties stated above in mind. Hence this is currently the most
plausible explanation for the presence of H$_2$ in the knots.

\acknowledgements

We thank William J. Henney and Robin J.R. Williams for fruitful discussions.
PvH, GVdS, KE, MG, JB, WDM, RH, CJ, SR, PR, and BV acknowledge support from
the Belgian Federal Science Policy Office via the PRODEX Programme of ESA.
PACS has been developed by a consortium of institutes led by MPE (Germany) and
including UVIE (Austria); KU Leuven, CSL, IMEC (Belgium); CEA, LAM (France);
MPIA (Germany); INAFIFSI/ OAA/OAP/OAT, LENS, SISSA (Italy); IAC (Spain). This
development has been supported by the funding agencies BMVIT (Austria),
ESA-PRODEX (Belgium), CEA/CNES (France), DLR (Germany), ASI/INAF (Italy), and
CICYT/MCYT (Spain). SPIRE has been developed by a consortium of institutes led
by Cardiff Univ. (UK) and including Univ. Lethbridge (Canada); NAOC (China);
CEA, LAM (France); IFSI, Univ. Padua (Italy); IAC (Spain); Stockholm
Observatory (Sweden); Imperial College London, RAL, UCL-MSSL, UKATC, Univ.
Sussex (UK); Caltech, JPL, NHSC, Univ. Colorado (USA). This development has
been supported by national funding agencies: CSA (Canada); NAOC (China); CEA,
CNES, CNRS (France); ASI (Italy); MCINN (Spain); SNSB (Sweden); STFC (UK); and
NASA (USA). Data presented in this paper were analysed using ``HIPE'', a joint
development by the Herschel Science Ground Segment Consortium, consisting of
ESA, the NASA Herschel Science Center, and the HIFI, PACS and SPIRE consortia.
This research made use of tools provided by Astrometry.net.

\bibliographystyle{aa}
\bibliography{14590}

\Online

\appendix

\section{Data reduction}
\label{reduction}

NGC 6720 was observed by PACS in scan map mode (one scan and one cross scan),
obtaining maps in the blue (70$\mu$m) and red (160$\mu$m) bands. The data were
processed with the Herschel Interactive data Processing Environment
\citep[HIPE,][]{Ott10}, following the recommended pipeline for these data.
This will be explained in detail in later papers, but a few deviations and
customizations are noted here. The WCS has an offset of a few arcseconds that
was corrected for the comparison of the PACS map to others (using known
sources in the field). We used the version 3 flatfield calibration. The
signals were converted to Jy using the version 5 calibration table. We did not
remove the cross-talk or corrected for response drifts. Cross-talk correction
is not part of the pipeline at present. Response drifts are unlikely to be a
problem for our observations: we find no drift in the calibration source
signals of more than 1.5\%\ over the duration of the observations. Removal of
the glitches (cosmic rays) was done in two stages: first from the regions
around our source by filtering along the time-dimension, and then from the
source itself working this time on the image plane. Cleaning the maps of the
1/f noise (the name refers to the type of power spectral density the noise
has) was done using a high-pass filter method. Here a filter passes over the
data as a function of time, subtracting the median of the data over a
specified time span (filter width). This allows the high noise frequencies to
pass and attenuates the lower frequencies. Considering that moving forward in
time means moving along a spatial direction as the instrument is scanning over
the source, it is clear that setting a low value for the filter width will
rapidly remove varying noise but can have the consequence of removing extended
emission. We processed the data with a variety of filter widths, but for our
source it made little difference to the photometry or morphology. Also worth
noting is that the high-pass filter creates artificial negative "sidelobes"
around strong sources. To deal with this the source was masked out before
running the high-pass filter. Finally the scan and cross scan frames were
combined and turned into a map.

Prior to making the maps presented in this paper the background sources were
removed with the HIPE sourceExtractorDaophot task. Slight WCS offsets for the
PACS and SPIRE maps were corrected, using background source coordinates
measured in the Calar Alto image. The pixel sizes and beam widths are given in
Table~\ref{phot}.

We custom-processed the raw Spitzer data obtained from the archive using both
the GeRT and Mosaicker software to remove some obvious anomalies apparent in the
pipeline-processed images by following the procedure explained by \citet{Ue06}.

We did the observations of NGC 6720 in the H$_2$ band by alternating object
and sky exposures. We reduced the data in the PixInsight Core package using
the acquired bias, dark, flatfield and sky frames. Special care was taken to
avoid artifacts in the object frames caused by the presence of stars in the
sky frames. A percentile clipping integration was done on the sky frames,
grouping them in groups of 5 images. By fitting the average signal of the
(already bias, dark and flatfield corrected) sky frames, a tight rejection of
outliers was possible, allowing the removal of the stars in the images. Then
the resulting sky frame was subtracted from the object frame acquired in the
middle of the five sky frames. We corrected the residual background gradients
due to sky variation by subtracting a sky background model built with the
DynamicBackgroundExtraction module of PixInsight. Finally this image was
astrometrically calibrated using the astrometry.net package \citep{La09}.

\section{Photometry}
\label{photometry}

\begin{table}
  \caption{Aperture photometry of NGC 6720 in various photometric bands. The apertures
    are elliptical and the semimajor axis is listed in the second column. The
    fluxes are listed with their measurement and calibration uncertainties, respectively.
    For the Spitzer data no measurement uncertainties were obtained. The AKARI data were
    taken from Izumiura et al.\ (in prep.). The pixel and beam sizes in the Herschel and
    Spitzer maps are also given.\label{phot}}
\begin{tabular}{lllllll}
\hline
$\lambda$ & aperture & flux & pixel size & beam size & source \\
$\mu$m & \arcsec & Jy & \arcsec & \arcsec \\
\hline
12 & & $0.82\pm0.07$ & & & IRAS \\
25 & & $10.1\pm0.5$ & & & IRAS \\
60 & & $52.\pm6.$ & & & IRAS \\
65 & & $55.\pm5.$ & & & AKARI \\
70  & 150 & $59.\pm7.\pm6.$ & 1 & $\sim5.2$ & PACS  \\
90 & & $70.\pm7.$ & & & AKARI \\
100 & & $55.\pm5.$ & & & IRAS \\
140 & & $37.\pm4.$ & & & AKARI \\
160 & 150 & $30.\pm3.\pm6.$ & 2 & $\sim12$ & PACS  \\
160 & 152  & $28.\pm7.$  & 8 & $\sim40$& Spitzer\\
160 & & $37.\pm14.$ & & & AKARI \\
250 & 168 & $13.\pm6.\pm4.$ & 6    & 18.1& SPIRE \\
350 & 150 & $8.\pm4.\pm2.$ & 10 & 25.2& SPIRE \\
500 & 154 &  $2.2\pm1.5\pm0.7$ & 14 & 36.6& SPIRE \\
\hline
\end{tabular}
\end{table} 

To measure the fluxes from the PACS, SPIRE and Spitzer maps we used an
elliptical aperture around NGC\,6720. The measured values are given in
Table~\ref{phot}. The PACS data were already in Jy/pix, the SPIRE data were
converted from Jy/beam to Jy/pix using the conversion $\pi({\rm
  beam/pixel})^2/(4\ln2)$ where the pixel and beam sizes are listed in
Table~\ref{phot}. The measured fluxes were also multiplied by factors provided
by the SPIRE team as the calibration tables did not yet include
these\footnote{The SPIRE flux calibration multipliers are: 250~$\mu$m: 1.02,
  350~$\mu$m: 1.05, 500~$\mu$m: 0.94.}. The Spitzer data were in MJy/sr and
were converted to Jy/pix via the conversion given in the MIPS instrument
handbook (Sect.~4.3). No additional corrections were applied (e.g. color
corrections). Uncertainties are quoted in Table~\ref{phot}. Calibration
uncertainties were taken from the respective instrument guides or release
notes. Measurement errors are difficult to calculate, as for these (bolometer)
instruments the Poissonian errors are not easily obtained. We combined the
contribution of the sky noise, the values in the error arrays which the PACS
and SPIRE pipelines create, and the variation in map fluxes that different
reasonable pipeline parameter variations gave. For the Spitzer fluxes no
measurement uncertainties were calculated.

As the beam size and the pixel scales on the maps are all different we
measured all the flux that could be seen from the source down to the sky
level, independently for each map. The aperture sizes used are included in
Table~\ref{phot}.

\section{The photoionization model}
\label{model}

We used the method described in \citet{vH99} to create an optimized
photoionization model of NGC 6720 using a prerelease of version C10.00 of the
photoionization code Cloudy (revision 3862). The method was slightly modified
in that we used \mbox{H-Ni} model atmospheres from \citet{Ra03}. As input we
used the UV and optical spectrum listed in \citet{Liu04}. We rejected the SWS
spectrum as it turned out that the aperture was mainly pointed at the central
``cavity'' and was therefore strongly biased towards the high excitation
region of the nebula. This made aperture correction factors highly dependent
on the ionization stage and hence very uncertain. We did use a re-reduced
version of the LWS spectrum of NGC 6720, adopting the aperture correction
factor listed in \citet{Liu04}. The [O\,{\sc i}] lines were excluded from the
modeling because they were fitted very badly. A plausible explanation is that
these lines are predominantly formed in the knots. We added two dust continuum
flux measurements from the LWS spectrum at 43 and 115~$\mu$m of
$4\times10^{-18}$ and $5\times10^{-19}$~W\,cm$^{-2}$\,$\mu$m$^{-1}$
respectively (aperture correction factors have not yet been applied to these
values). We also used two radio continuum flux densities at 4850 and 1400~MHz
of 360 and 440~mJy respectively, which are averages of the data collected by
\citet{Vo09}. For the angular diameter we used 76\arcsec\ \citep{Liu04}. We
adopted a distance of 740~pc \citep{OD09}. The model was stopped when the
electron temperature dropped below 4000~K. The resulting optimized model has
the parameters listed in Table~\ref{model:par}. Most symbols have their usual
meaning. $\Gamma$ denotes the dust-to-gas mass ratio and $\epsilon$ the
logarithmic abundance of an element ($\epsilon$(H) $\equiv$ 12.00). The
electron temperature and density are averaged over the ionized nebula. We will
refer to this model as the standard Cloudy model of NGC~6720.

\end{document}